\documentclass[twocolumn,english,prb,epsfig,rotate,showpacs,aps,longbibliography]{revtex4-2}
\usepackage[T1]{fontenc}
\usepackage[latin9]{inputenc}
\usepackage{color}
\usepackage{verbatim}
\usepackage{amsmath}
\usepackage{amssymb}
\usepackage{graphicx}
\usepackage{esint}
\usepackage{microtype}
\usepackage{hyperref}
\usepackage{lmodern}
\makeatletter
\usepackage{babel}
\usepackage{titlesec}
\usepackage{epsfig}
\usepackage{amsmath}
\usepackage{graphicx}
\usepackage{dcolumn}
\usepackage{bm}
\usepackage{amsfonts,amssymb}
\usepackage[T1]{fontenc}
\usepackage[latin9]{inputenc}
\usepackage{amsmath}
\usepackage{graphicx}
\usepackage{amssymb}
\usepackage{esint}
\usepackage{babel}
\usepackage{microtype}
\usepackage{xcolor}
\usepackage[normalem]{ulem}
\newcommand{\blue}{\color{black}}

\begin{document}

\renewcommand{\figurename}{Fig.}
\title{Non-Hermitian quantum geometric tensor and nonlinear electrical response} 
\author{Kai Chen$^{\chi}$}
\author{Jie Zhu$^{+}$}
\affiliation{School of Physics Science and Engineering, Tongji University, Shanghai, China}

\date{\today}

\begin{abstract}
We demonstrate that the non-Hermitian quantum geometric tensor (QGT) governs nonlinear electrical responses in systems with a spectral line gap. The quantum metric, which is the symmetric component of the QGT and takes complex values in non-Hermitian systems, generates an intrinsic nonlinear conductivity independent of the scattering time. In contrast, the full complex-valued QGT induces a distinct conductivity that depends explicitly on the wavepacket width. Using one- and two-dimensional non-Hermitian models, we establish a direct link between nonlinear dynamics and the QGT, thereby connecting quantum state geometry to observable transport phenomena. Crucially, we reveal that the finite wavepacket width fundamentally alters non-Hermitian transport---a mechanism strictly absent in Hermitian systems. This framework elucidates non-Hermitian response theory by revealing how the complex geometry of quantum states, captured by the QGT, and the wavepacket width jointly encode transport in open and synthetic quantum matter. 
\end{abstract}
\maketitle

\section{Introduction}
Higher-order responses play a crucial role in condensed matter physics, as they often reveal connections to emerging physical concepts. One prominent example is the quantum geometric tensor (QGT), which characterizes the geometric structure of quantum states in projective Hilbert space~\cite{provost1980riemannian,kolodrubetz2017geometry,rossi2021quantum} and influences a system's nonlinear response to external driving fields. The QGT consists of two key components: its antisymmetric part, known as the Berry curvature~\cite{berry1984quantal}, and its symmetric part, the Fubini-Study quantum metric (QM)~\cite{fubini1904sulle,study1905kurzeste}. In non-Hermitian systems, both the Berry curvature and the QM can be complex-valued. These geometric quantities underpin a vast array of phenomena, with the Berry curvature central to effects such as the quantum anomalous Hall effect~\cite{liu2016quantum,chang2023colloquium}, topological insulators and superconductors~\cite{hasan2010colloquium,qi2011topological}, Weyl and Dirac semimetals~\cite{young2012dirac,hosur2013recent,young2015dirac,yan2017topological}, orbital magnetization~\cite{thonhauser2005orbital,resta2010electrical,xiao2010berry}, and the converse vortical effect~\cite{chen2024chiral}. {\blue{The quantum metric, in contrast, captures the distance between neighboring quantum states in Hilbert space. A nonzero quantum metric implies that quantum states change nontrivially as parameters vary; the larger the metric, the more distinguishable the states become for a fixed parameter shift.}} In Hermitian systems, the semipositive definite nature of the QGT~\cite{provost1980riemannian} imposes constraints on the trace and determinant of its real and imaginary parts, leading to important physical inequalities~\cite{roy2014band,rossi2021quantum}.

The significance of the QGT has been further amplified by recent experimental advances. Interest in the QGT has surged, driven by its critical influence on nonlinear material responses and the discovery of flat-band superconductivity in twisted moir{\'e} systems \cite{kim2017tunable,cao2018correlated,cao2018unconventional,yankowitz2019tuning,chen2019signatures,lu2019superconductors}. Since then, numerous QGT-related phenomena have been (re)discovered, further expanding the field. For instance, the QM has been found to contribute to superfluid weight \cite{peotta2015superfluidity,xie2020topology,liang2017band,peri2021fragile,chen2025effect}, optical conductivity \cite{ghosh2024probing,ezawa2024analytic}, nonlinear conductivity \cite{ezawa2024intrinsic,das2023intrinsic}, nonlinear transport \cite{wang2023quantum,fang2024quantum,mandal2024quantum} and the nonlinear quantum valley Hall effect \cite{das2024nonlinear}. This expanding landscape demonstrates that the QGT is not merely a formal tool but a central actor in determining diverse material responses.

While the QGT framework is well-established in Hermitian physics, its extension to non-Hermitian systems reveals a host of new phenomena. Non-Hermitian systems exhibit remarkable properties that distinguish them from their Hermitian counterparts~\cite{bender2007making,el2018non}. A defining feature is the skin effect \cite{bergholtz2021exceptional,zhang2022review}, where eigenstates localize exponentially at boundaries due to extreme sensitivity to boundary conditions, fundamentally altering spectral behavior. Another hallmark is the emergence of non-Hermitian exceptional points \cite{ding2022non,kawabata2019classification}, spectral singularities at which eigenvalues and eigenvectors coalesce, collapsing the Hilbert space and enabling exotic phase transitions. Non-Hermiticity also modifies foundational frameworks such as linear response theory and transport phenomena \cite{pan2020non,sticlet2022kubo,wang2022anomalous}, leading to counterintuitive effects absent in Hermitian systems. These phenomena, including unidirectional invisibility and enhanced sensitivity, lack direct analogs in Hermitian physics, underscoring the unique theoretical and experimental implications of non-Hermitian theories.  

A crucial distinction in non-Hermitian systems arises from the difference between complex conjugation and transposition, which makes their symmetry structure more intricate \cite{kawabata2019symmetry,chen2022non}. This added complexity extends the 10-fold way topological classification \cite{altland1997nonstandard}, originally formulated for Hermitian systems, into a broader 38-fold Bernard-LeClair class that accounts for non-Hermitian symmetries \cite{zhou2019periodic,kawabata2019symmetry,okuma2023non}. Furthermore, the complex nature of the energy spectrum necessitates a refined definition of band gaps. Unlike in Hermitian systems, non-Hermitian band gaps can be categorized as point or line gaps \cite{ashida2020non,kawabata2019classification}, both of which play a crucial role in the topological classification of non-Hermitian phases.

Given this rich non-Hermitian landscape, a natural question arises: how do quantum geometric properties manifest and influence physical responses? The geometric properties of non-Hermitian systems have attracted significant attention. {\blue{The concept of a geometric phase in non-Hermitian systems was first introduced by Garrison and Wright~\cite{garrison1988complex}, and later formalized in the context of topological band theory by Shen \textit{et al.}~\cite{shen2018topological}. Building on these foundations, recent works have explored the non-Hermitian QGT, revealing its impact on anomalous wavepacket dynamics and transport phenomena~\cite{silberstein2020berry, wang2022anomalous, hu2024generalized, hu2025quantum, behrends2025quantum}. }}However, these studies generally rely on the zero-width wavepacket approximation ($W \rightarrow 0$), thereby disregarding vital corrections arising from the finite spatial extent of the electron wavepacket. While this approximation is often sufficient in Hermitian systems, we find that in the non-Hermitian regime, the finite wavepacket width couples to the imaginary parts of the complex Berry curvature and quantum metric. This coupling leads to significant corrections in the nonlinear electrical response that are explicitly dependent on $W$, a feature that has previously been overlooked.

In this work, we bridge this gap by investigating the role of the full non-Hermitian QGT, including finite-wavepacket effects, in the nonlinear response of line-gapped systems. We show that although line-gapped non-Hermitian systems are topologically equivalent to Hermitian systems~\cite{schindler2023hermitian}---as they can be adiabatically connected to a Hermitian limit---they exhibit unique nonlinear responses absent in Hermitian systems. This distinction stems from the complex structure of the QGT in non-Hermitian systems, which fundamentally alters the mechanism of nonlinear conductivity compared to Hermitian cases. We demonstrate that the complex QM generates an intrinsic nonlinear conductivity, while the full QGT induces a distinct second-order DC conductivity (SODCc) component proportional to the wavepacket width---a feature exclusive to non-Hermitian systems. Furthermore, we analyze subsystems faithfully described by non-Hermitian Hamiltonians, which emerge due to interactions with their environment. As a paradigmatic example, we compute the nonlinear surface conductivity of a Chern insulator and isolate the distinct contributions of the QGT to this response. Our findings underscore the critical role of quantum geometric structures in non-Hermitian systems, revealing novel physical properties and nonlinear transport phenomena with no Hermitian analogs.

\vspace{-10pt} 
\section{Nonlinear DC conductivity and QGT in non-Hermitian systems}

A fundamental distinction between non-Hermitian and Hermitian systems is that, in non-Hermitian systems, the left and right eigenstates of the Hamiltonian are different. These eigenstates satisfy the biorthogonality condition $\langle\psi^{\mathrm{L}}_{m,\mathbf{k}}|\psi^{\mathrm{R}}_{n,\mathbf{k}}\rangle = \delta_{mn}$, where $\mathbf{k}$ denotes the Bloch momentum and $m,n$ label the band indices. Specifically, for a non-Hermitian Hamiltonian $H_{\mathrm{NH}}(\mathbf{k})$, the right eigenstates $|\psi^{\mathrm{R}}_{i,\mathbf{k}}\rangle$ and left eigenstates $\langle\psi^{\mathrm{L}}_{i,\mathbf{k}}|$ satisfy the eigenvalue equations $H_{\mathrm{NH}}(\mathbf{k})|\psi^{\mathrm{R}}_{i,\mathbf{k}}\rangle = \xi_{i,\mathbf{k}}|\psi^{\mathrm{R}}_{i,\mathbf{k}}\rangle$ and $\langle\psi^{\mathrm{L}}_{i,\mathbf{k}}| H_{\mathrm{NH}}(\mathbf{k}) = \xi_{i,\mathbf{k}}\langle\psi^{\mathrm{L}}_{i,\mathbf{k}}|$, where $\xi_{i,\mathbf{k}} \in \mathbb{C}$ is the complex energy, and resolve the identity operator as $I = \sum_i |\psi^{\mathrm{R}}_{i,\mathbf{k}}\rangle\langle\psi^{\mathrm{L}}_{i,\mathbf{k}}|$. Motivated by this distinction, we define operator expectations in the left-right basis as $\langle \psi^L\mid\hat{O}\mid\psi^R\rangle$ for any operator $\hat{O}$.

{\blue{
Given this biorthogonal structure, a consistent definition of geometric quantities is required. We explicitly define the non-Hermitian QGT as:
\begin{equation}
    T_{\mu\nu} = \langle \partial_{\mu} \psi_n^L | \partial_{\nu} \psi_n^R \rangle - \langle \partial_{\mu} \psi_n^L | \psi_n^R \rangle \langle \psi_n^L | \partial_{\nu} \psi_n^R \rangle.
\end{equation}
In this non-Hermitian framework, the symmetric component corresponds to the non-Hermitian Quantum Metric, while the antisymmetric component corresponds to the non-Hermitian Berry curvature. We acknowledge that there are multiple generalizations of the QGT in the literature, often leading to different physical conclusions depending on the context~\cite{hu2024generalized, hu2025quantum}. We adopt the biorthogonal (``Left-Right'') definition here because it ensures gauge invariance~\cite{silberstein2020berry, hu2024generalized} and is consistent with the framework required to describe anomalous geometric transport~\cite{wang2022anomalous} and gauge-invariant wavepacket spreading~\cite{hu2024generalized, hu2025quantum}.}}

Having established the geometric framework, we now develop the dynamical theory to connect the QGT to transport. We investigate how the non-Hermitian QGT governs the nonlinear DC conductivity. To establish this connection, we adopt a semiclassical framework, where the current density is given by $\mathbf{J} = -e \int_{\mathbf{k}} f \dot{\mathbf{r}}$, with \(\int_{\mathbf{k}}\equiv \int\frac{d^D k}{(2\pi)^D}\) denoting momentum-space integration in a \(D\)-dimensional system. Here, \(\dot{\mathbf{r}}\) represents the semiclassical electron velocity, and \(f\) is the distribution function.

The generalized semiclassical equations of motion for a non-Hermitian system under an external electric field \(\mathbf{E}\) are given by~\cite{dong2025non}:
\begin{align}
    \dot{\mathbf{r}} &= \operatorname{Re}\!\left[ \nabla_{\mathbf{k}}\xi_{n,\mathbf{k}} + e\mathbf{E} \times \boldsymbol{\Omega}_{n,\mathbf{k}} \right], \label{eq:eom1} \\
    \dot{\mathbf{k}} &= -e\mathbf{E} + W^{2}\, \operatorname{Im}\!\left[ \nabla_{\mathbf{k}}\xi_{n,\mathbf{k}} + e\mathbf{E} \times \boldsymbol{\Omega}_{n,\mathbf{k}} \right], \label{eq:eom2}
\end{align}
where \(e\) is the electron charge, \(W\) denotes the Gaussian wave-packet width, and \(\operatorname{Re}(\cdot)\) and \(\operatorname{Im}(\cdot)\) extract the real and imaginary parts, respectively. The non-Hermitian Berry curvature is defined as $\boldsymbol{\Omega}_{n,\mathbf{k}} \equiv i\nabla_{\mathbf{k}} \times \langle\psi^{\mathrm{L}}_{n,\mathbf{k}}|\nabla_{\mathbf{k}}\psi^{\mathrm{R}}_{n,\mathbf{k}}\rangle$.

{\blue{The explicit dependence on $W$ is a unique non-Hermitian effect arising from the gradient of the imaginary energy spectrum, $\text{Im}(\xi_{k})$, and complex Berry curvature. A finite-width wavepacket samples this gain/loss landscape, leading to preferential amplification and a net drift of the center-of-mass towards high-gain regions~\cite{dong2025non}. This ``geometric drift,'' absent in Hermitian dynamics, is the physical origin of the $W^2$ scaling in the nonlinear conductivity.}}

{\blue{We note that the theoretical derivation of nonlinear response functions is currently a subject of active discussion. Generally, three distinct frameworks are employed: semiclassical wavepacket dynamics, the Luttinger-Kohn thermodynamic approach, and quantum kinetics based on the density matrix \cite{qiang2025clarification}. Recent studies have highlighted that these methods can occasionally yield contradictory results in the nonlinear regime, particularly regarding the precise prefactors of geometric contributions \cite{qiang2025clarification}. In this work, we adopt the semiclassical wavepacket approach. We choose this framework because it offers a transparent physical intuition connecting transport to the QGT and, crucially, allows for the explicit incorporation of the finite wavepacket width $W$. This enables us to isolate the width-dependent contributions to the conductivity, which are less accessible in the standard thermodynamic limit assumed by the Luttinger-Kohn formalism.}}

With the dynamics established, we now apply this framework to calculate the nonlinear conductivity. In this section, we first focus on the limit of zero wave-packet width (\(W \to 0\)) to isolate the intrinsic geometric contributions. (The influence of a finite wave-packet width on the SODCc will be examined in the following section.)

We analyze the nonlinear response of the system to a uniform electric field, which enters the Hamiltonian via the minimal coupling as a perturbative potential \(-e\mathbf{E} \cdot \mathbf{r}\). {{\blue{To derive the second-order nonlinear response, we employ the Boltzmann transport equation under the relaxation time approximation:

\vspace{-0.4cm}
\begin{equation}
\begin{aligned}
    \partial_{t}f + \dot{\mathbf{k}}\cdot\nabla_{\mathbf{k}}f = -\frac{f-f_{0}}{\tau},
\end{aligned}
\end{equation}
where $f$ is the distribution function, $f_0$ is the equilibrium distribution, and $\tau$ is the scattering time. In the DC limit, we expand the distribution function in powers of the electric field $\mathbf{E}$ as $f = f_{0} + f^{(1)} + f^{(2)} + \dots$. In the limit of zero wavepacket width, the semiclassical dynamics are governed by $\dot{\mathbf{k}} = -e\mathbf{E}$, leading to the recursive solution $f^{(n)} = \tau e \mathbf{E} \cdot \nabla_{\mathbf{k}} f^{(n-1)}$.

Crucially, the external electric field not only drives the electron distribution but also modifies the band structure itself through interband mixing. To capture this, we apply the Schrieffer-Wolff transformation to the non-Hermitian Hamiltonian $H_{NH} = H_0 - e\mathbf{E}\cdot\mathbf{r}$. This perturbative approach reveals field-induced corrections to the band energy $\xi_{\mathbf{k}}$ and the Berry curvature $\mathbf{\Omega}_{\mathbf{k}}$:
\begin{align}
    \xi_{\mathbf{k}} &\approx \xi_{\mathbf{k}}^{(0)} + \xi_{\mathbf{k}}^{(2)} = \xi_{\mathbf{k}}^{(0)} + e^2 G_{\mu\nu}^{LR} E^\mu E^\nu, \\
    \Omega_{\mathbf{k}} &\approx \Omega_{\mathbf{k}}^{(0)} + \Omega_{\mathbf{k}}^{(1)},
\end{align}
where $G_{\mu\nu}^{LR}$ is the band-renormalized non-Hermitian QM derived from the second-order energy correction, and $\Omega_{\mathbf{k}}^{(1)}$ represents the first-order field correction to the Berry curvature.

The total current density is calculated by integrating the velocity over the Brillouin zone: $\mathbf{J} = -e \int_{\mathbf{k}} f \dot{\mathbf{r}}$. By combining the expanded distribution function ($f^{(0)}, f^{(1)}, f^{(2)}$) with the field-corrected semiclassical velocity \(\dot{\mathbf{r}} = \text{Re}[\nabla_{\mathbf{k}}\xi_{\mathbf{k}} + e\mathbf{E}\times\boldsymbol{\Omega}_{\mathbf{k}}]\), we identify three distinct contributions to the SODCc: a Drude-like term proportional to $\tau^2$, a Berry curvature dipole term proportional to $\tau$, and an intrinsic geometric term independent of $\tau$. Summing these components yields the total SODCc:}}

\begin{equation}
\begin{aligned}
    \sigma_{\theta\mu\nu} &= -\int_{\mathbf{k}} \bigg\{ \tau^2 \text{Re}\left(\partial_\theta \xi^{(0)}_{\mathbf{k}}\right) \partial_{\mu}\partial_{\nu} f_0 \\
    &\quad -\tau f_0 \frac{\epsilon_{\theta\mu l}\partial_{\nu} \text{Re}\Omega_{l}^{(0)} + \epsilon_{\theta\nu l}\partial_{\mu} \text{Re}\Omega_{l}^{(0)}}{2}  \\
       &\quad+ f_{0}\text{Re}\left( 2\partial_\theta G^{LR}_{\mu\nu} - \frac{\partial_{\nu}G_{\mu \theta}^{LR} + \partial_{\mu}G_{\nu \theta}^{LR}}{2} \right) \bigg\},
\end{aligned}
\label{sigma22}
\end{equation}
where we suppress the band index to simplify the notation, the indices $\theta$, $\mu$, and $\nu$ label the spatial directions, and $\epsilon_{\mu\nu\theta}$ is the antisymmetric Levi-Civita symbol, and the Einstein summation convention is implied for repeated spatial indices. The equilibrium distribution takes the explicit form $f_0 = \left( 1 + \exp(\text{Re}\,\xi^{(0)}_{\mathbf{k}} / k_B T) \right)^{-1}$. {\blue{Throughout this work, we set the reduced Planck constant $\hbar=1$, the Boltzmann constant $k_B = 1$, and the electron charge $e = 1$. Consequently, the scattering time $\tau$ is measured in units of the inverse hopping amplitude.}}

{\blue{The third term in Eq.~(\ref{sigma22}) is independent of the scattering time and is proportional to the band-renormalized non-Hermitian QM, denoted as $G_{\mu\nu}^{LR}$. Unlike the conventional Fubini-Study metric which characterizes distances in the projective Hilbert space, this quantity arises directly from the second-order perturbation to the band energy induced by the interband Berry connection. It is defined as:
\begin{equation}
    G_{\mu\nu}^{LR} \equiv G_{n,\mu\nu}^{LR} = \sum_{m\ne n}\frac{A_{nm,\mu}^{LR(0)}A_{mn,\nu}^{LR(0)}+A_{nm,\nu}^{LR(0)}A_{mn,\mu}^{LR(0)}}{2(\xi_{n,k}^{(0)}-\xi_{m,k}^{(0)})},
\end{equation}
where $n$ denotes the target band index. Here, the unperturbed non-Hermitian Berry connection is given by $A^{LR(0)}_{mn,\mu} \equiv i \langle \psi^{L(0)}_{m,\mathbf{k}} \mid \partial_{\mu} \psi^{R(0)}_{n,\mathbf{k}} \rangle$, and the unperturbed non-Hermitian Berry curvature is $\boldsymbol{\Omega}^{(0)} \equiv i \nabla_{\mathbf{k}} \times \mathbf{A}^{LR(0)}_{nn}$. We derive this expression via the Schrieffer-Wolff transformation (see Appendix A), where the second-order energy perturbation $\xi_{n,k}^{(2)}$ relates to the external field bilinearly as $\xi_{n,k}^{(2)}=e^{2}G_{\mu\nu}^{LR}E^{\mu}E^{\nu}$. Consequently, the denominator contains the energy difference to the first power, $(\xi_{n,k}^{(0)}-\xi_{m,k}^{(0)})$, which is characteristic of static second-order perturbation theory, distinguishing it from the squared energy difference typically found in optical transition elements.}}

{\blue{Notably, the analytical structure of Eq.~(\ref{sigma22}) formally mirrors the result derived for Hermitian systems~\cite{kaplan2024unification}. Specifically, the dependence of the nonlinear conductivity on the gradients of the band dispersion, Berry curvature, and QM is algebraically identical to the Hermitian case. The crucial distinction in the non-Hermitian context is that these geometric quantities are intrinsically complex-valued. Consequently, the real-part operations in Eq.~(\ref{sigma22}) are necessary to project these complex geometric contributions onto the physically observable, real-valued current, thereby highlighting the unique influence of non-Hermitian effects.}}

\section{Effect of Wavepacket Width on SODCc.}

In the previous section, we derived the SODCc in the limit of zero wave-packet width. We now extend the analysis to incorporate a finite wave-packet width, which introduces novel contributions governed by the complex structure of the QGT. Assuming the imaginary component of the band dispersion is perturbatively small compared to all other energy scales (e.g., real band energies and scattering rates), we derive the SODCc as (see Appendix A for details):
\vspace{-1.5ex}
\begin{equation}
\sigma_{\mu\nu\theta}^{W}=\sigma_{\mu\nu\theta}^{W=0}+\int_{\mathbf{k}} W^2\tau \Gamma_{2,1}+ W^2 \tau^2\Gamma_{2,2}+ W^4\tau^2\Gamma_{4,2},
\label{full}
\end{equation}
\vspace{-\baselineskip}

The term $\sigma_{\mu\nu\theta}^{W=0}$ represents the DC conductivity in the zero wavepacket width limit, i.e., Eq.~(\ref{sigma22}). The coefficients $\Gamma_{2,1}$, $\Gamma_{2,2}$, and $\Gamma_{4,2}$ encode the interplay between finite wavepacket width and the complex QGT. These terms are defined as
\begin{equation}
\begin{aligned}
\Gamma_{2,1} &=   \Bigg[ \frac{\epsilon_{i\nu l} \epsilon_{\mu\theta\gamma} \partial_i f_0 \, \mathrm{Im}\Omega_{l}^{(0)} \, \mathrm{Re}\Omega_{\gamma}^{(0)} + (\theta \leftrightarrow \nu)}{2} \\
&\quad + \frac{\left(2\partial_{\theta} \mathrm{Im}G_{\nu i}^{LR} - \partial_i \mathrm{Im}G_{\nu\theta}^{LR}\right) \partial_{i}f_{0} + (\theta \leftrightarrow \nu)}{2} \Bigg] \partial_{\mu} \mathrm{Re}\xi_{\mathbf{k}}^{(0)}, \\
\Gamma_{2,2} &=\frac{1}{2} \Bigg[ \epsilon_{j\theta i} \Bigl( \partial_{\nu} \mathrm{Im}\,\Omega_{i}^{(0)} \partial_j f_0 \\
&\quad + 2 \, \mathrm{Im}\,\Omega_{i}^{(0)} \partial_j \partial_{\nu}f_0 \Bigr) + (\theta \leftrightarrow \nu) \Bigg] \partial_{\mu} \mathrm{Re}\,\xi_{\mathbf{k}}^{(0)},\\
\Gamma_{4,2} &= \frac{\epsilon_{i\nu j} \epsilon_{n\theta\gamma} \, \mathrm{Im}\Omega^{(0)}_j \, \partial_i \left( \mathrm{Im}\Omega^{(0)}_{\gamma} \partial_n f_0 \right) + (\theta \leftrightarrow \nu)}{2} \, \partial_{\mu} \mathrm{Re}\xi_{\mathbf{k}}^{(0)}.
\end{aligned}
\end{equation}

Having presented the general result, we can now analyze its geometric content. The coefficients $\Gamma_{2,2}$ and $\Gamma_{4,2}$ depend solely on the imaginary part of the non-Hermitian Berry curvature. In contrast, the coefficient $\Gamma_{2,1}$ involves contributions from both the full complex non-Hermitian Berry curvature and the imaginary part of the non-Hermitian band-renormalized QM.
This structure reveals a key physical insight: the finite wavepacket width couples directly to the geometric quantities. This coupling generates distinct conductivity contributions scaling with $W^2$ and $W^4$, which are unique signatures of non-Hermitian quantum geometry and have no counterpart in Hermitian systems.

Having established the theoretical result, we now address its experimental implications. A natural question concerns the experimental relevance of the wavepacket width $W$, which governs the magnitude of the geometric nonlinear response (scaling as $W^2$). In solid-state quantum materials, effective non-Hermiticity naturally arises from finite quasiparticle lifetimes or coupling to an environment. In this regime, the effective spatial extent of the wavepacket $W$ is intrinsic and is typically determined by the carrier coherence length. Semiclassically, this width is physically associated with the thermal de Broglie wavelength, $\lambda_{th} \propto T^{-1/2}$~\cite{xiao2010berry, Ashcroft1976}.

Consequently, the parameter $W$ is naturally tunable via temperature. This suggests that the geometric contributions to the nonlinear conductivity (proportional to $W^2$) will exhibit a distinct $T$ scaling. This specific temperature dependence allows the geometric non-Hermitian response to be experimentally disentangled from extrinsic scattering-induced mechanisms. Conversely, this relationship implies that nonlinear transport measurements can serve as a diagnostic tool: by isolating the $T$ component of the nonlinear conductivity, one can experimentally extract the effective coherence length of the non-Hermitian system.

\section{SODCc in one-dimensional non-Hermitian models.}

To illustrate how the QGT manifests in a non-Hermitian system's SODCc, in this section we consider two concrete models within the zero-width wavepacket approximation: a paradigmatic one-dimensional lattice model and a more complex boundary system. The effect of a finite wavepacket width on the SODCc will be examined for a two-dimensional model in the next section.
\vspace{-20pt} 
\subsection{Non-Hermitian Su-Schrieffer-Heeger model.}

\begin{figure}[h]
\includegraphics[width=1\columnwidth,height=1.2\textheight,keepaspectratio]{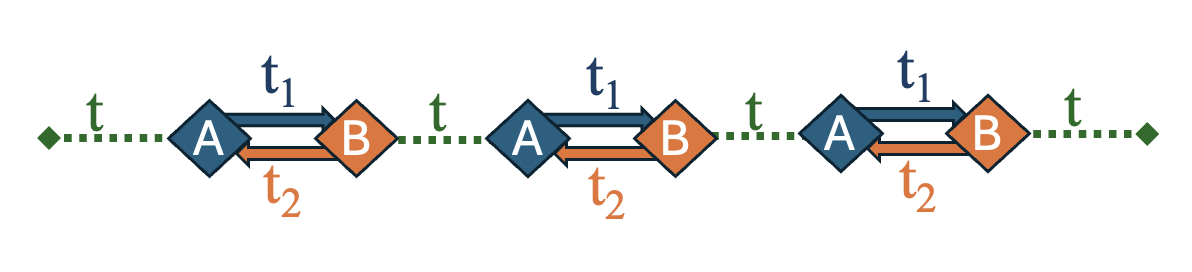}
\caption{(Color online) Schematic of the non-Hermitian SSH lattice model.}
\label{figNHSSH} 
\end{figure}

We begin with the non-Hermitian Su-Schrieffer-Heeger (SSH) model, a one-dimensional topological insulator featuring nonreciprocal intracell hopping \cite{chen2024quantum}. The lattice structure is schematically depicted in Fig.~\ref{figNHSSH}. The Bloch Hamiltonian is given by
\begin{equation}
\begin{aligned}
H_{\text{nHSSH}}(k) &= \left(\frac{t_1+t_2}{2}+t\cos k\right)\sigma_x \\
&\quad + \left(i\frac{t_1-t_2}{2}+t\sin k\right)\sigma_y,
\end{aligned}
\end{equation}
where $\sigma_i$ ($i = x, y$) are the Pauli matrices, $t \in \mathbb{R}$ is the intercell hopping, and $t_1 \in \mathbb{C}$ and $t_2 \in \mathbb{R}$ are the intracell directional hoppings. The Hamiltonian becomes Hermitian when $t_1 = t_2\in \mathbb{R}$.

For this one-dimensional model, the Berry curvature vanishes. Consequently, the general expression for the SODCc in Eq.~(\ref{sigma22}) simplifies significantly to:
\begin{equation}
\begin{aligned}
\sigma_{xxx}=-\text{Re} \int_{\mathbf{k}} f_0 \left\{\tau^2\partial^3_x \xi^{(0)}_{\mathbf{k}}+ \partial_x G^{LR}_{xx}\right\}.
\end{aligned}
\label{sigma3}
\end{equation}
This simplified form isolates two contributions: a conventional Drude-like term ($\propto \tau^2$) and an intrinsic geometric term involving the gradient of the QM. The non-Hermitian SSH model preserves time-reversal symmetry, given by $H_{\text{nHSSH}}(k)^* = H_{\text{nHSSH}}(-k)$ when $t_1 \in \mathbb{R}$. Under this symmetry, the band dispersion satisfies $\xi^{(0)*}_{\mathbf{k}} = \xi^{(0)}_{-\mathbf{k}}$, and the Berry connection obeys $A^{LR(0)}_{mn,\nu}(k)^* = A^{LR(0)}_{mn,\nu}(-k)$, which forces $\sigma_{xxx} = 0$. Therefore, we introduce a complex $t_1$ to break time-reversal symmetry, while carefully choosing its value to maintain a finite line gap [Fig.~\ref{fig1}(a)].

We now analyze the resulting conductivity. As shown in Eq.~(\ref{sigma3}), the real part of the band-renormalized QM provides an intrinsic contribution to the SODCc, which is independent of $\tau$. Its temperature dependence is shown in Fig.~\ref{fig1}(b). The ratio of the intrinsic term $\sigma_{xxx}^{0}$ to the total SODCc $\sigma_{xxx}$ increases with $T$. In the strong scattering regime ($\tau = 1$), the intrinsic term dominates the SODCc [Fig.~\ref{fig1}(c)]. In contrast, in the weak scattering regime ($\tau = 10$), the intrinsic contribution is small at low $T$ but increases with temperature.

\begin{figure}[h]
\includegraphics[width=1\columnwidth,height=1.2\textheight,keepaspectratio]{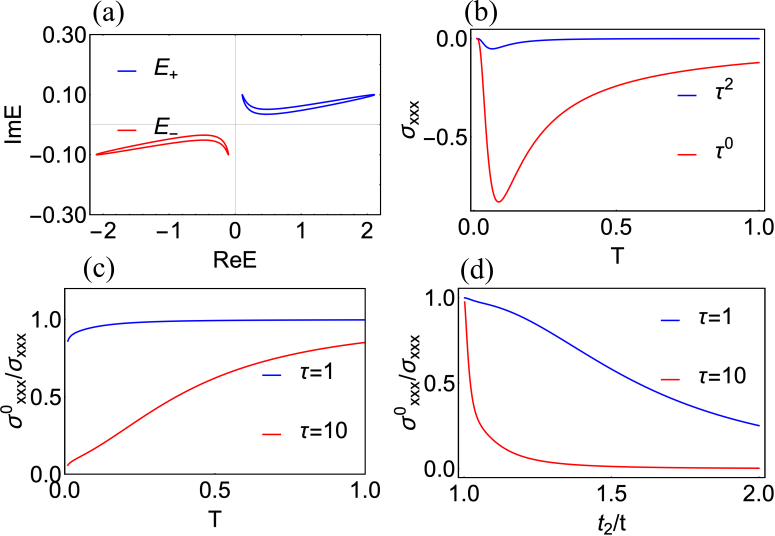}
\caption{SODCc in the non-Hermitian SSH model. (a) Band dispersion for \(t_2/t = 1.1\). (b) Intrinsic contribution (order \(\tau^0\)) and the coefficient of the \(\tau^2\) term in the SODCc. (c) Ratio of the intrinsic contribution to the total SODCc as a function of \(T\) for \(t_2/t = 1.1\); (d) the same ratio as a function of \(t_2/t\) for \(T = 0.1\). Other parameters: \(t_1/t = t_2/t + i\,0.2\).}
\label{fig1}
\end{figure}

The behavior near the topological phase transition is particularly revealing. The energy gap of the non-Hermitian SSH model closes at the critical point $t_2/t = 1$. Near this critical value, the intrinsic contribution dominates the SODCc in both the strong ($\tau = 1$) and weak ($\tau = 10$) scattering regimes [Fig.~\ref{fig1}(d)]. This demonstrates that quantum geometric effects become particularly pronounced near topological phase transitions, where the band-renormalized quantum metric becomes singular.

\subsection{Boundary of a Chern insulator.}

Having examined a minimal lattice model, we now consider a system with a more intricate physical origin: the effective boundary of a Chern insulator. This example illustrates how non-Hermitian Hamiltonians emerge naturally in subsystems coupled to an environment. Previous studies \cite{hamanaka2024non} have shown that the low-energy physics of this boundary can be effectively captured by a non-Hermitian Hamiltonian, where non-Hermiticity arises from the coupling between the boundary and the bulk, thereby inducing a complex self-energy. The effective non-Hermitian Hamiltonian takes the form
\begin{equation}
\begin{aligned}
    H_{edge}(k) &= \left(\sin{k} - \frac{1-\left(m+\cos{k}\right)^2}{2\left(i\eta-\sin{k}\right)}\right)\sigma_y + \left(m+\cos{k}\right)\sigma_z \\
    &\quad + \frac{1-\left(m+\cos{k}\right)^2}{2\left(i\eta-\sin{k}\right)}\sigma_0.
\end{aligned}
\label{chern}
\end{equation}

This model also exhibits a phase with a line gap [see Fig.~\ref{fig2}(a)]. We compute its SODCc to test the generality of the geometric effects found in the SSH model. As illustrated in Fig.~\ref{fig2}(b), the intrinsic contribution to the total SODCc increases with temperature in both the strong and weak scattering regimes. At low temperatures, similar to the non-Hermitian SSH model, the intrinsic contribution is more pronounced in the strong scattering regime than in the weak scattering regime.

\begin{figure}[h]
\includegraphics[width=1\columnwidth,height=1.2\textheight,keepaspectratio]{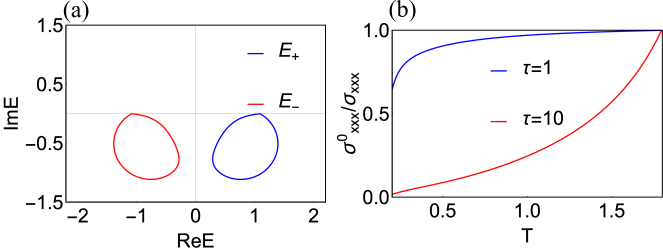}
\caption{SODCc at the boundary of a Chern insulator. (a) Band dispersion of the Hamiltonian in Eq.~(\ref{chern}). (b) Ratio of the intrinsic contribution to the total SODCc as a function of temperature $T$. The parameters are $m = -1.9$ and $\eta = 1/\sqrt{80}$.}
\label{fig2}
\end{figure}

Together, these two examples illustrate the applicability of our framework to distinct physical settings. The non-Hermitian SSH model shows how intrinsic geometric conductivity arises in a simple, tunable lattice. The Chern insulator boundary model confirms that these effects persist in more complex, physically motivated systems where non-Hermiticity emerges from environmental coupling. In both cases, the QGT provides the fundamental link between the complex band structure and the measurable nonlinear response.

\begin{figure}[h]
\includegraphics[width=1\columnwidth,height=1.2\textheight,keepaspectratio]{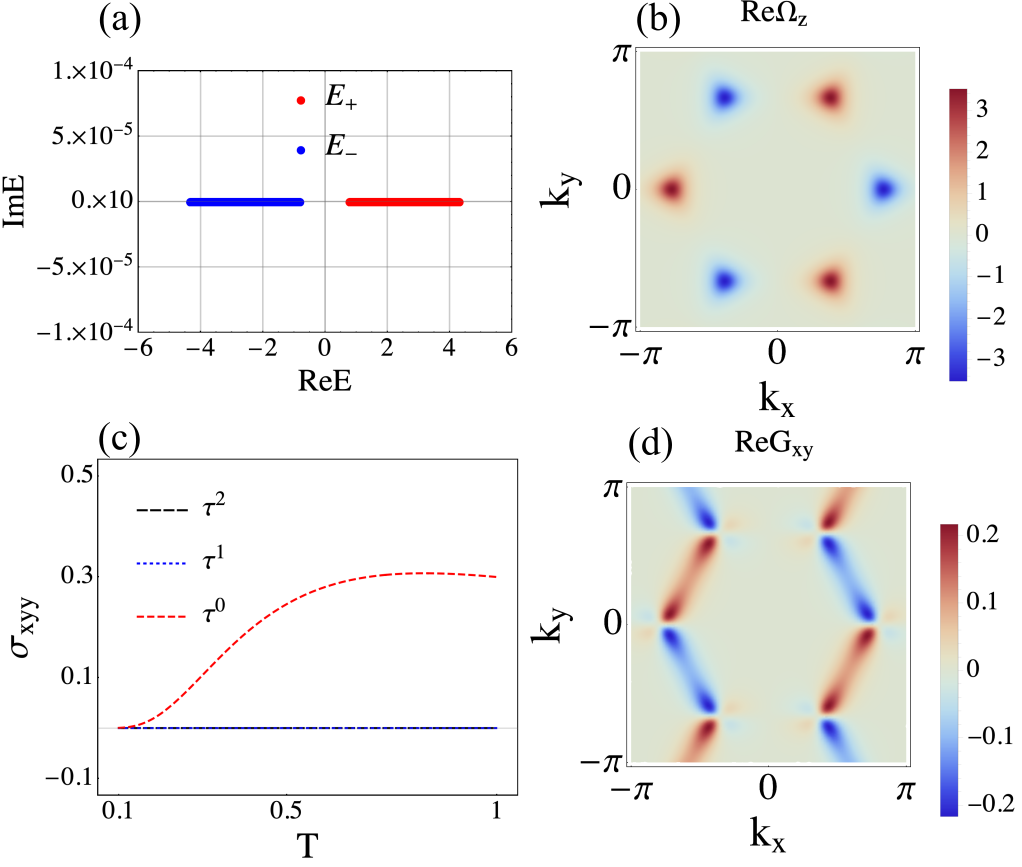}
\caption{Band dispersion (a), (c) SODCc $\sigma_{xyy}$ as a function of $T$, and momentum-space distributions of the Berry curvature (b) and the $xy$-component of the QM, $G_{xy}$ (d), for parameters $m = 0.8$, $\gamma = 0.5$.}
\label{fig3} 
\end{figure}

\section{SODCc in a two-dimensional non-Hermitian model and the effect of wavepacket width.}

We now extend our analysis to a two-dimensional system to investigate the effect of a finite wavepacket width on the SODCc. To isolate this effect, we first examine a model with a real energy spectrum, where the QGT components are purely real. We then introduce a controlled perturbation to generate complex eigenvalues and complex QGT components, enabling us to study the direct coupling between wavepacket width and the imaginary parts of the geometry.

We consider a two-dimensional non-Hermitian model described by the Hamiltonian \cite{long2022non}:
\begin{equation}
\begin{aligned}
H_{2d}(k_x, k_y) =
\begin{bmatrix}
m & h\left(k_x, k_y\right) \\
\frac{1}{\gamma}h\left(k_x, k_y\right)^*& -m
\end{bmatrix}
\end{aligned}
\label{real}
\end{equation}
where the function $h(k_x,k_y)=e^{-ik_y}+2e^{ik_y/2}\cos\left(\frac{\sqrt{3}}{2}k_x\right)$
defines the system describing a non-reciprocal lattice such that for $\gamma\neq 1$
the Hamiltonian is non-Hermitian yet can exhibit real eigenvalues [Fig.~\ref{fig3}(a)].
Note that the reality of the spectrum in Eq.~\ref{real} arises from its pseudo-Hermitian construction \cite{long2022non}, which guarantees real eigenvalues for all $\gamma > 0$, distinct from $\mathcal{PT}$-symmetric systems that generally undergo symmetry breaking.

{\color{black} The real eigenvalues of the Hamiltonian allow us to disentangle the
effects of the non-Hermitian QGT from the dissipative or
amplifying effects typically associated with complex eigenvalues.} Furthermore, both
the Berry curvature and the QM are real-valued functions of the Bloch momentum
$\mathbf{k}$; accordingly, Figs.~\ref{fig3}(b) and \ref{fig3}(d) display the
distributions of the Berry curvature $\text{Re}\Omega_z(k_x,k_y)$ and the QM component
$\text{Re}G_{xy}$ in $\mathbf{k}$-space. Notably, the Berry curvature satisfies $\Omega_z(k_x,k_y)=-\Omega_z(-k_x,k_y)$ and $\Omega_z(k_x,k_y)=\Omega_z(k_x,-k_y)$, while the QM component obeys $G_{xy}(k_x,k_y)=-G_{xy}(-k_x,k_y)$ and $G_{xy}(k_x,k_y)=-G_{xy}(k_x,-k_y)$. Taking the $xyy$-component of the SODCc, $\sigma_{xyy}$, as a concrete example and noting that $E_{\pm}(\mathbf{k})=E_{\pm}(-\mathbf{k})$, both the first- and second-order $\tau$ terms in Eq.~(\ref{sigma3}) vanish, so that only the band-renormalized QM, which is independent of $\tau$, contributes to $\sigma_{xyy}$ [see Fig.~\ref{fig3}(c)].

Having established the behavior for the purely real-spectrum case, we now examine the effect of finite wavepacket width on the SODCc. In non-Hermitian systems, the QGT possesses a complex structure. Whereas the imaginary part of the QGT in Hermitian systems corresponds to the purely real Berry curvature, the Berry curvature in non-Hermitian systems becomes intrinsically complex. This complexity introduces an imaginary component that couples dynamically to the spatial width of quantum wavepackets via the semiclassical equation $\dot{\mathbf{k}} = -e\mathbf{E} + W^2\,\text{Im}\!\left[\nabla_{\mathbf{k}}\xi_{n,\mathbf{k}} + e\mathbf{E} \times \boldsymbol{\Omega}_{n,\mathbf{k}}\right] $. The conventional zero-width approximation suppresses this coupling by neglecting finite-$W$ contributions. To reveal the role of this coupling in the nonlinear response, we introduce a small imaginary perturbation to the parameter $m$ in the Hamiltonian~(\ref{real}). This yields a system with complex eigenvalues and a line gap [Fig.~\ref{fig4}(a)], while the QM and Berry curvature acquire a complex character [Figs.~\ref{fig4}(c--f)]. As a result, the coefficients $\Gamma_{i,j}$ become nonzero, and a finite wavepacket width produces a pronounced enhancement of $\sigma_{xxx}^{W}$, as shown in Fig.~\ref{fig4}(b).

\begin{figure}[h]
\includegraphics[width=1\columnwidth,height=1.2\textheight,keepaspectratio]{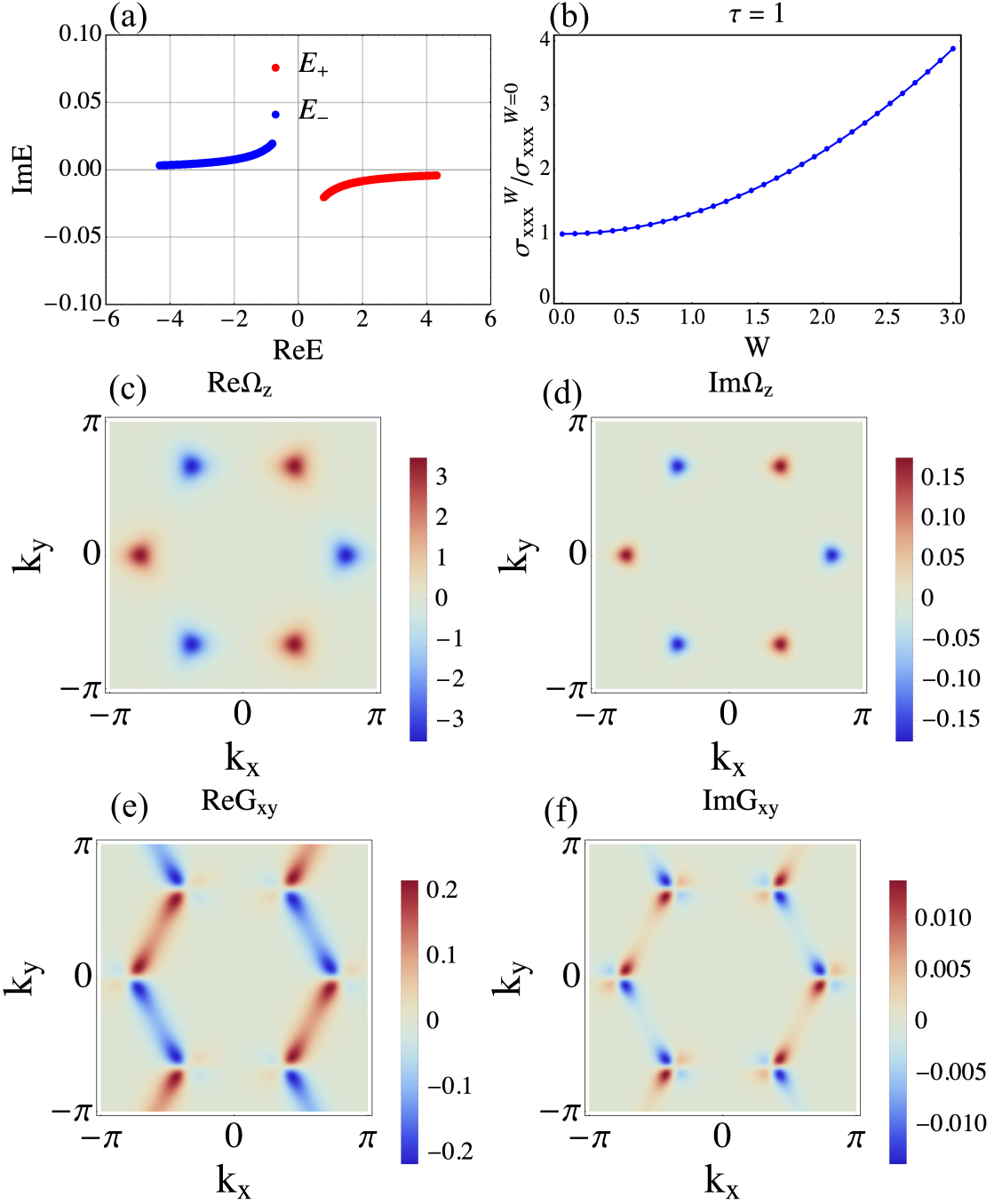}
\caption{Band dispersion (a), SODCc $\sigma_{xxx}$ as a function of $W$ (b), and momentum-space distributions of typical components of the QGT (c-f). Parameters: $m = 0.8 - 0.02i$, $\gamma = 0.5$.}
\label{fig4} 
\end{figure}

\section{Conclusions and Outlook.}
In this work, we have elucidated the role of the QGT in governing nonlinear electrical responses in non-Hermitian systems with a spectral line gap. In the zero-wavepacket-width limit, the SODCc depends on the real part of the band dispersion, the Berry curvature, and the band-renormalized QM. Beyond this limit, the finite wavepacket width bridges the complex-valued band-renormalized QM and Berry curvature---unique features of the projected Hilbert space in non-Hermitian systems---to the nonlinear response. Using one- and two-dimensional non-Hermitian models, we have demonstrated the distinct role of the QM in shaping the SODCc. Notably, in the zero-width regime, our results indicate that the band-renormalized QM dominates the SODCc at higher temperatures and under conditions of strong scattering (short $\tau$). Furthermore, a finite wavepacket width amplifies the SODCc, a phenomenon exclusive to non-Hermitian systems that arises from the imaginary components of the band dispersion and Berry curvature.

Our findings extend naturally to surfaces and edges of topological materials, such as Chern insulators, which are often effectively described by non-Hermitian Hamiltonians. This connection enables the prediction of surface nonlinear conductivity in otherwise Hermitian systems. By bridging quantum geometry with nonlinear transport, this work establishes a versatile framework for harnessing geometric effects in topological devices and engineered materials.

\acknowledgments

\textit{Acknowledgments}---K.C. acknowledges support from the Shanghai Magnolia (Yulan) Pu River Project (Grant No. 24PJD121) and from the Fundamental Research Funds for the Central Universities. J.Z. acknowledges support from the National Natural Science Foundation of China (Grants No. 92263208), the National Key R\&D Program of China (Grants No. 2022YFA1404400), and the Research Grants Council of Hong Kong SAR (Grant No. AoE/P-502/20).

\noindent $^{\chi}$: KaiChenPhys@tongji.edu.cn
\noindent $^{+}$: jiezhu@tongji.edu.cn

\bibliography{nonHer}

 \onecolumngrid

\appendix 
\section{Nonlinear electric response and the quantum geometric tensor}

In this section, we derive the nonlinear electric response of a non-Hermitian system under an external electric field and reveal its connection to the quantum geometric tensor associated with the system. Starting from the semiclassical equation of motion \cite{dong2025non}:

\begin{equation} \left\{ \begin{aligned}
    \dot{\mathbf{r}}&= \text{Re}\left[\nabla_{\mathbf{k}}\xi_{n,\mathbf{k}} +e\mathbf{E}\times \boldsymbol{\Omega_{n,\mathbf{k}}}\right], \\
    \dot{\mathbf{k}}&=-e\mathbf{E}+W^2 \text{Im}\left[\nabla_{\mathbf{k}}\xi_{n,\mathbf{k}} +e\mathbf{E}\times \boldsymbol{\Omega_{n,\mathbf{k}}}\right]\equiv \mathbf{W}_{\mathbf{k}}.
\end{aligned} \right. \label{eom0}\end{equation}
where $\mathbf{E}$ is the external electric field, $W$ is the width of the Gaussian wavepacket, and $\xi_{n,\mathbf{k}}$ and $\boldsymbol{\Omega}_{n,\mathbf{k}}\equiv i \nabla_{\mathbf{k}} \times \langle\psi^{L}_{n,\mathbf{k}}\mid \nabla_{\mathbf{k}}\psi^{R}_{n,\mathbf{k}}\rangle$ denote the $n$th band energy and the Berry curvature, respectively.  

In this work, we focus on a nondegenerate band and, to simplify the notation when there is no risk of confusion, omit the target band index. Under the relaxation time approximation \cite{wang2022anomalous} and assuming a spatially uniform distribution function \(f\), the Boltzmann equation for the electron distribution is given by:

\begin{equation} 
 \partial_t f +\dot{\mathbf{k}}\cdot \nabla_{\mathbf{k}}f = -\frac{f-f_0}{\tau},
\label{Boltzmann}
\end{equation}
where \(\tau\) is the scattering time and \(f_0\) is the equilibrium distribution function in the absence of external fields. In a non-Hermitian system, for the target band, we have
\[
f_0 = \frac{1}{1+\exp\left(\mathrm{Re}\,\xi_{\mathbf{k}}/k_B T\right)}.
\]

The electron current is given by:
\begin{equation} 
 \mathbf{J}=-e \int_{\mathbf{k}} f \dot{\mathbf{r}}=-e \int_{\mathbf{k}} f \mathbf{v}.
 \label{jj}
\end{equation}
The higher-order response is extracted via the Schrieffer-Wolff transformation, as demonstrated in Refs.~\cite{kaplan2024unification,fang2024quantum} for Hermitian systems. In the weak-field limit, we expand the distribution function \(f\), the group velocity \(\mathbf{v}\), the band dispersion \(\xi_{\mathbf{k}}\), and the Berry curvature \(\boldsymbol{\Omega}_{\mathbf{k}}\) in powers of the external field \(\mathbf{E}\) as follows:

\begin{equation} \left\{ \begin{aligned}
    f&=\sum_{n=0} f^{\left(n\right)}, \mathbf{v}=\sum_{n=0} \mathbf{v}^{\left(n\right)}, \\
    \xi_{\mathbf{k}}&= \sum_{n=0}\xi_{\mathbf{k}}^{\left(n\right)}, \boldsymbol{\Omega}_{\mathbf{k}}= \sum_{n=0}\boldsymbol{\Omega}^{\left(n\right)}_{\mathbf{k}}, \\
    \mathbf{W}_{\mathbf{k}}&= \sum_{n=0} \mathbf{W}^{\left(n\right)}_{\mathbf{k}}, \mathbf{J}= \sum_{n=0} \mathbf{J}^{\left(n\right)}_{\mathbf{k}}.
\end{aligned} \right. 
\label{expand0}
\end{equation}

From the first equation in Eqs.~(\ref{eom0}), the three lowest-order terms in the expansion of the group velocity are given by:

\begin{equation} \left\{ \begin{aligned}
    \mathbf{v}^{\left(0\right)}&=\text{Re}\nabla_{\mathbf{k}}\xi^{(0)}_{\mathbf{k}}, \\
    \mathbf{v}^{\left(1\right)}&=\text{Re}\left[\nabla_{\mathbf{k}}\xi^{(1)}_{\mathbf{k}}+e\mathbf{E}\times\boldsymbol{\Omega}^{(0)}_\mathbf{k}\right], \\
    \mathbf{v}^{\left(2\right)}&=\text{Re}\left[\nabla_{\mathbf{k}}\xi^{(2)}_{\mathbf{k}}+e\mathbf{E}\times\boldsymbol{\Omega}^{(1)}_\mathbf{k}\right],  \\
    ...
\end{aligned} \right. \label{expandv}\end{equation}

Using the ansatz \(f = f\, e^{i\omega t}\) and the second equation in Eqs.~(\ref{eom0}), Eq.~(\ref{Boltzmann}) becomes

\begin{equation}
    i\omega f +\mathbf{W}_{\mathbf{k}}\cdot \nabla_{\mathbf{k}}f = -\frac{f-f_0}{\tau}
\end{equation}
Substituting Eqs.~(\ref{expand0}) into the above equation, we obtain:
\begin{equation}
    \sum_{n=0}i\omega f^{\left(n\right)} +\sum_{i,j=0}\mathbf{W}^{\left(i\right)}_{\mathbf{k}}\cdot \nabla_{\mathbf{k}}f^{\left(j\right)} = \frac{f_0-f^{(0)}}{\tau}-\sum_{l=1} \frac{f^{\left(l\right)}}{\tau}
\end{equation}
Then $f^{(n)}$ can be obtained recursively as:

\begin{equation} \left\{ \begin{aligned}
    i\omega f^{(0)} +\mathbf{W}^{\left(0\right)}_{\mathbf{k}}\cdot \nabla_{\mathbf{k}}f^{\left(0\right)}&=\frac{f_0-f^{(0)}}{\tau}, \\
    i\omega f^{(1)} + \mathbf{W}^{\left(0\right)}_{\mathbf{k}}\cdot \nabla_{\mathbf{k}}f^{\left(1\right)}+\mathbf{W}^{\left(1\right)}_{\mathbf{k}}\cdot \nabla_{\mathbf{k}}f^{\left(0\right)}&=-\frac{f^{(1)}}{\tau}, \\
    i\omega f^{(2)} +\mathbf{W}^{\left(0\right)}_{\mathbf{k}}\cdot \nabla_{\mathbf{k}}f^{\left(2\right)}+\mathbf{W}^{\left(2\right)}_{\mathbf{k}}\cdot \nabla_{\mathbf{k}}f^{\left(0\right)}+\mathbf{W}^{\left(1\right)}_{\mathbf{k}}\cdot \nabla_{\mathbf{k}}f^{\left(1\right)}&=-\frac{f^{(2)}}{\tau},  \\
    ...
\end{aligned} \right. \label{expand}\end{equation}
where $\mathbf{W}^{\left(0\right)}_{\mathbf{k}}=W^2\nabla_{\mathbf{k}} \text{Im}\xi^{(0)}_{\mathbf{k}}$, $\mathbf{W}^{\left(1\right)}_{\mathbf{k}}=-e\mathbf{E}+W^2 \text{Im}\left[\nabla_{\mathbf{k}}\xi^{(1)}_{\mathbf{k}}+e\mathbf{E}\times \boldsymbol{\Omega}^{(0)}_{\mathbf{k}}\right]$ and $\mathbf{W}^{\left(m\geq 2\right)}_{\mathbf{k}}=W^2 \text{Im}\left[\nabla_{\mathbf{k}}\xi^{(m)}_{\mathbf{k}}+e\mathbf{E}\times \boldsymbol{\Omega}^{(m-1)}_{\mathbf{k}}\right]$.
In the direct current limit, where $\omega \ll 1/\tau$, the above equations simplify to:
\begin{equation} \left\{ \begin{aligned}
    \mathbf{W}^{\left(0\right)}_{\mathbf{k}}\cdot \nabla_{\mathbf{k}}f^{\left(0\right)}&=\frac{f_0-f^{(0)}}{\tau}, \\
     \mathbf{W}^{\left(0\right)}_{\mathbf{k}}\cdot \nabla_{\mathbf{k}}f^{\left(1\right)}+\mathbf{W}^{\left(1\right)}_{\mathbf{k}}\cdot \nabla_{\mathbf{k}}f^{\left(0\right)}&=-\frac{f^{(1)}}{\tau}, \\
    \mathbf{W}^{\left(0\right)}_{\mathbf{k}}\cdot \nabla_{\mathbf{k}}f^{\left(2\right)}+\mathbf{W}^{\left(2\right)}_{\mathbf{k}}\cdot \nabla_{\mathbf{k}}f^{\left(0\right)}+\mathbf{W}^{\left(1\right)}_{\mathbf{k}}\cdot \nabla_{\mathbf{k}}f^{\left(1\right)}&=-\frac{f^{(2)}}{\tau},  \\
    ...
\end{aligned} \right. \label{fn}\end{equation}

Substituting the solutions for \(f^{(n)}\) and those from Eqs.~(\ref{expandv}) into Eq.~(\ref{jj}) yields, in principle, the electrical response to any order.

\subsection{zero wavepacket width}
In this subsection, we consider the limit of a zero wave packet width, such that the term proportional to \( W^2 \) in \( \mathbf{W}_{\mathbf{k}} \) can be safely neglected. We then obtain:

\begin{equation} \left\{ \begin{aligned}
    f^{(0)}&=f_0, \\
     f^{(1)}&=\tau e\mathbf{E}\cdot \nabla_{\mathbf{k}}f^{\left(0\right)}, \\
   f^{(2)}&=\tau e\mathbf{E}\cdot \nabla_{\mathbf{k}}f^{\left(1\right)},  \\
    ...
\end{aligned} \right. \end{equation}

The currents to first and second order in the electric field are given by:

\begin{equation} \left\{ \begin{aligned}
    \mathbf{J}^{(1)}&=-e\int_{\mathbf{k}}f^{(1)}\text{Re}\nabla_{\mathbf{k}}\xi^{(0)}_{\mathbf{k}}+f^{(0)}\text{Re}\left[\nabla_{\mathbf{k}}\xi^{(1)}_{\mathbf{k}}+e\mathbf{E}\times\boldsymbol{\Omega}^{(0)}_\mathbf{k}\right], \\
     &=-e\int_{\mathbf{k}}\tau e\mathbf{E}\cdot \nabla_{\mathbf{k}}f^{\left(0\right)}\text{Re}\nabla_{\mathbf{k}}\xi^{(0)}_{\mathbf{k}}+f^{(0)}\text{Re}\left[\nabla_{\mathbf{k}}\xi^{(1)}_{\mathbf{k}}+e\mathbf{E}\times\boldsymbol{\Omega}^{(0)}_\mathbf{k}\right],
\end{aligned} \right. \end{equation}
and
\begin{equation} 
    \mathbf{J}^{(2)}=-e\int_{\mathbf{k}}f^{(0)}\text{Re}\left[\nabla_{\mathbf{k}}\xi^{(2)}_{\mathbf{k}}+e\mathbf{E}\times\boldsymbol{\Omega}^{(1)}_\mathbf{k}\right]+f^{(1)}\text{Re}\left[\nabla_{\mathbf{k}}\xi^{(1)}_{\mathbf{k}}+e\mathbf{E}\times\boldsymbol{\Omega}^{(0)}_\mathbf{k}\right] +f^{(2)}\text{Re}\nabla_{\mathbf{k}}\xi^{(0)}_{\mathbf{k}} 
\end{equation}

To analyze the corrections to the band dispersion and Berry curvature induced by an external field \( \mathbf{E} \), we expand \( \xi_{\mathbf{k}} \) and \( \boldsymbol{\Omega}_{\mathbf{k}} \) up to third order in \( \mathbf{E} \). In the presence of the external field, the system Hamiltonian takes the form  

\begin{equation}
    H_{NH}=\sum_{m,n}\left(\xi^{(0)}_{n,\mathbf{k}}\delta_{mn}-e\langle\psi^{L}_{m,\mathbf{k}}\mid\mathbf{E}\cdot\mathbf{r}\mid\psi^{R}_{n,\mathbf{k}}\rangle\right)\mid\psi^{R}_{m,\mathbf{k}}\rangle\langle\psi^{L}_{n,\mathbf{k}}\mid \equiv H_0 + H_1
\end{equation}
with
\begin{equation} \left\{ \begin{aligned}
    H_0&=\sum_{n}\left(\xi^{(0)}_{n,\mathbf{k}}-e\mathbf{E}\cdot\mathbf{A}^{LR}_{nn}\right)\mid\psi^{R}_{n,\mathbf{k}}\rangle\langle\psi^{L}_{n,\mathbf{k}}\mid, \\
     H_1&=\sum_{m\neq n}-e\mathbf{E}\cdot\mathbf{A}^{LR}_{mn}\mid\psi^{R}_{m,\mathbf{k}}\rangle\langle\psi^{L}_{n,\mathbf{k}}\mid,
\end{aligned} \right. \end{equation}
where $\mathbf{A}^{LR}_{mn}\equiv i \langle\psi^{L}_{m,\mathbf{k}}\mid \nabla_{\mathbf{k}}\psi^{R}_{n,\mathbf{k}}\rangle$. 
The Schrieffer-Wolff operator $S$ perturbs an operator $O$, yielding a transformed operator $O' = e^{S} O e^{-S}$. 

The operator $S$ is chosen to satisfy $H_1 + [S, H_0] = 0$ which leads to 
\begin{equation}
    \sum_{m\neq n}e\mathbf{E}\cdot\mathbf{A}^{LR}_{mn}\mid\psi^{R}_{m,\mathbf{k}}\rangle\langle\psi^{L}_{n,\mathbf{k}}\mid=\sum_{i,j}S_{ij}\mid\psi^{R}_{i,\mathbf{k}}\rangle\langle\psi^{L}_{j,\mathbf{k}}\mid\left[\left(\xi^{(0)}_{j,\mathbf{k}}-e\mathbf{E}\cdot\mathbf{A}^{LR}_{jj}\right)-\left(\xi^{(0)}_{i,\mathbf{k}}-e\mathbf{E}\cdot\mathbf{A}^{LR}_{ii}\right)\right],
\end{equation}
we have:
\begin{equation} \left\{ \begin{aligned}
    S_{nn}&=0,\\
    S_{m\neq n}&=\frac{e\mathbf{E}\cdot\mathbf{A}^{LR}_{mn}}{\left(\xi^{(0)}_{n,\mathbf{k}}-e\mathbf{E}\cdot\mathbf{A}^{LR}_{nn}\right)-\left(\xi^{(0)}_{m,\mathbf{k}}-e\mathbf{E}\cdot\mathbf{A}^{LR}_{mm}\right)}\approx \frac{e\mathbf{E}\cdot\mathbf{A}^{LR}_{mn}}{\left(\xi^{(0)}_{n,\mathbf{k}}-\xi^{(0)}_{m,\mathbf{k}}\right)}+\frac{e\mathbf{E}\cdot\mathbf{A}^{LR}_{mn}\left(e\mathbf{E}\cdot\mathbf{A}^{LR}_{nn}-e\mathbf{E}\cdot\mathbf{A}^{LR}_{mm}\right)}{\left(\xi^{(0)}_{n,\mathbf{k}}-\xi^{(0)}_{m,\mathbf{k}}\right)^2}+O(E^3).
\end{aligned} \right. \end{equation}
where, in the second equation, we use the fact that there exists a gap between the target band and the remaining bands.

The Hamiltonian is transform to

\begin{equation}
    H_{NH}^{'}=e^{S}H_{NH}e^{-S}=H_0 +\frac{\left[S,H_1\right]}{2}+\frac{\left[S,\left[S,H_1\right]\right]}{3}+...,
\end{equation}
where

\begin{equation}
    \frac{\left[S,H_1\right] }{2}= \frac{-e}{2}\sum_{m,n}\left(\sum_{i\neq n}S_{mi}\mathbf{E}\cdot\mathbf{A}^{LR}_{in}-\sum_{i\neq m}\mathbf{E}\cdot\mathbf{A}^{LR}_{mi}S_{in}\right)\mid\psi^{R}_{m,\mathbf{k}}\rangle\langle\psi^{L}_{n,\mathbf{k}}\mid\equiv\frac{-e}{2}\sum_{m, n}F_{mn}\mid\psi^{R}_{m,\mathbf{k}}\rangle\langle\psi^{L}_{n,\mathbf{k}}\mid,
\end{equation}
and 
\begin{equation}
    \frac{\left[S,\left[S,H_1\right]\right]}{3}=O(E^3)
\end{equation}
Now, the perturbed band dispersion $ \xi_{n,\mathbf{k}}=\langle\psi^{L}_{n,\mathbf{k}}\mid H^{'}\mid\psi^{R}_{n,\mathbf{k}}\rangle$, up to second order in the electric field, is given by:
\begin{equation} \left\{ \begin{aligned}
    \xi^{(1)}_{n,\mathbf{k}}&=-e\mathbf{E}\cdot\mathbf{A}^{LR}_{nn}=0,\\
    \xi^{(2)}_{n,\mathbf{k}}&=\frac{-e^2}{2}\left(\sum_{m\neq n} \frac{\mathbf{E}\cdot\mathbf{A}^{LR}_{mn}\mathbf{E}\cdot\mathbf{A}^{LR}_{nm}+\mathbf{E}\cdot\mathbf{A}^{LR}_{nm}\mathbf{E}\cdot\mathbf{A}^{LR}_{mn}}{\left(\xi^{(0)}_{m,\mathbf{k}}-\xi^{(0)}_{n,\mathbf{k}}\right)}\right)\equiv e^2\sum_{\mu,\nu} G^{LR}_{n,\mu\nu}E^{\mu}E^{\nu}.
\end{aligned} \right. \end{equation}
where, in the first equation, we choose a gauge for the Berry curvature to make it perpendicular to the electric field. Here, $G_{n,\mu\nu}^{LR}=\sum_{i\neq n}\frac{A^{LR}_{ni,\mu}A^{LR}_{in,\nu}+A^{LR}_{ni,\nu}A^{LR}_{in,\mu}}{2\left(\xi^{(0)}_{n,\mathbf{k}}-\xi^{(0)}_{i,\mathbf{k}}\right)}$ denotes the band-renormalized non-Hermitian QM.

Similarly, the Berry connection $ \mathbf{A}^{LR} = \sum_{m,n} \mathbf{A}_{mn}^{LR} \mid\psi^{R}_{m,\mathbf{k}}\rangle \langle\psi^{L}_{n,\mathbf{k}}\mid $ transforms to:
\begin{equation}
    \mathbf{A}^{'LR}=\sum_{m,n}\mathbf{A}_{mn}^{LR}\mid\psi^{R}_{m,\mathbf{k}}\rangle\langle\psi^{L}_{n,\mathbf{k}}\mid+\sum_{m,n} \left(\sum_{j\neq m}\frac{e\mathbf{E}\cdot\mathbf{A}^{LR}_{mj}\mathbf{A}_{jn}^{LR}}{\left(\xi^{(0)}_{j,\mathbf{k}}-\xi^{(0)}_{m,\mathbf{k}}\right)}-\sum_{j\neq n}\frac{e\mathbf{A}_{mj}^{LR}\mathbf{E}\cdot\mathbf{A}^{LR}_{jn}}{\left(\xi^{(0)}_{n,\mathbf{k}}-\xi^{(0)}_{j,\mathbf{k}}\right)}\right)\mid\psi^{R}_{m,\mathbf{k}}\rangle\langle\psi^{L}_{n,\mathbf{k}}\mid+O(E^2)
\end{equation}
Therefore, the perturbed Berry connection for the $n$th band is given by:
\begin{equation}
    \mathbf{A}_{n}^{(1)LR}=-e\sum_{j\neq n}\frac{\mathbf{E}\cdot\mathbf{A}^{LR}_{nj}\mathbf{A}_{jn}^{LR}+\mathbf{A}_{nj}^{LR}\mathbf{E}\cdot\mathbf{A}^{LR}_{jn}}{\left(\xi^{(0)}_{n,\mathbf{k}}-\xi^{(0)}_{j,\mathbf{k}}\right)}=-e2G_{n,\mu\nu}^{LR}E^{\mu}\hat{x}_{\nu},
\end{equation}
where $\hat{x}_{\nu}$ denotes the unit vector in the $\nu$ direction, and the corresponding Berry curvature is:
\begin{equation}
    \Omega_{n,\alpha}^{(1)LR}=-2e\epsilon_{\alpha\beta\nu}\partial_{\beta}G_{n,\mu\nu}^{LR}E^{\mu},
\end{equation}

Combining the above results, we can express the response current up to second order in the electric field, as shown below:

\begin{equation} \left\{ \begin{aligned}
    \mathbf{J}^{(1)}
     &=-\int_{\mathbf{k}}\tau e^2\mathbf{E}\cdot \nabla_{\mathbf{k}}f^{\left(0\right)}\text{Re}\nabla_{\mathbf{k}}\xi^{(0)}_{\mathbf{k}}+f^{(0)}\text{Re}\left[e^2\mathbf{E}\times\boldsymbol{\Omega}^{(0)}_\mathbf{k}\right], \\
     J_{i}^{(1)}&=-E^{j}\int_{\mathbf{k}}\tau e^2 \partial_j f_0 \partial_{i}\text{Re}\xi_{\mathbf{k}}^{(0)}+e^2f_0 \epsilon_{ijl}\text{Re}\Omega_{l}^{(0)}=E^{j}\sigma_{ij},
\end{aligned} \right. \end{equation}
and
\begin{equation} \left\{ \begin{aligned} 
    \mathbf{J}^{(2)}&=-e\int_{\mathbf{k}}f^{(0)}\text{Re}\left[\nabla_{\mathbf{k}}\xi^{(2)}_{\mathbf{k}}+e\mathbf{E}\times\boldsymbol{\Omega}^{(1)}_\mathbf{k}\right]+f^{(1)}\text{Re}\left[e\mathbf{E}\times\boldsymbol{\Omega}^{(0)}_{\mathbf{k}}\right] +f^{(2)}\text{Re}\nabla_{\mathbf{k}}\xi^{(0)}_{\mathbf{k}}, \\
    J_{i}^{(2)}&=-e\int_{\mathbf{k}}f_{0}\text{Re}\left[\partial_i\xi^{(2)}_{\mathbf{k}}+e\epsilon_{ijl}E^{j}\Omega_{l}^{(1)}\right]+f^{(1)}\text{Re}\left[e\epsilon_{ijl}E^{j}\Omega_{l}^{(0)}\right] +f^{(2)}\text{Re}\partial_i \xi^{(0)}_{\mathbf{k}}=\sigma_{ijl}E^{j}E^l. 
\end{aligned} \right. \end{equation}

where
\begin{equation} \left\{ \begin{aligned}
    f^{(0)}&=f_0=\frac{1}{1 + \exp^{\text{Re}\xi_{\mathbf{k}}/k_B T}}, \\
     f^{(1)}&=\tau e\mathbf{E}\cdot \nabla_{\mathbf{k}}f^{\left(0\right)}=e\tau E^{j}\partial_j f_0, \\
   f^{(2)}&=\tau e\mathbf{E}\cdot \nabla_{\mathbf{k}}f^{\left(1\right)}=\left(e\tau\right)^2 E^{i}E^{j}\partial_i\partial_j f_0,  \\
 \Omega_{\alpha}^{(1)}&=-2e\epsilon_{\alpha\beta\nu}\partial_{\beta}G_{\mu\nu}^{LR}E^{\mu}, \\
 \xi_{\mathbf{k}}^{(2)}&=e^2 G^{LR}_{\mu\nu}E^{\mu}E^{\nu}
\end{aligned} \right. \end{equation}

\begin{equation}
\left\{ \begin{aligned} 
    \sigma_{i\mu\nu}=\frac{\partial^2 J_{i}^{(2)}}{2\partial E^{\mu}\partial E^{\nu}}&=-e^3\int_{\mathbf{k}}\left\{\tau^2 \partial_{\mu}\partial_{\nu} f_0\text{Re}\left(\partial_i \xi^{(0)}_{\mathbf{k}}\right)-\tau f_0 \frac{\epsilon_{i\mu l}\partial_{\nu} \text{Re}\Omega_{l}^{(0)}+\epsilon_{i\nu l}\partial_{\mu} \text{Re}\Omega_{l}^{(0)}}{2}+f_{0}\text{Re}\left[ 2\partial_iG^{LR}_{\mu\nu}-\frac{\partial_{\nu}G_{\mu i}^{LR}+\partial_{\mu}G_{\nu i}^{LR}}{2}\right]\right\}, \\
    &=\sigma_{i\nu\mu}
\end{aligned} \right. \label{sigma2}\end{equation}
Note that $G^{LR}_{\mu\nu}=G^{LR}_{\nu\mu}$.

\subsection{Effects of Wave Packet Width}  
In the previous subsection, we considered the limit of a zero wave packet, where the time evolution of momentum is solely dictated by the external electric field. However, in this limit, certain non-Hermitian effects are suppressed. Here, we extend our analysis to finite wave packet widths and examine their impact on the DC conductivity in non-Hermitian systems. To facilitate analytical progress, we focus on a non-Hermitian model in which the imaginary part of the spectrum varies slightly in \(\mathbf{k}\)-space within certain parameter regimes. In this case, the semiclassical equations of motion are modified as follows:

\begin{equation} \left\{ \begin{aligned}
    \dot{\mathbf{r}}&= \text{Re}\left[\nabla_{\mathbf{k}}\xi_{n,\mathbf{k}} +e\mathbf{E}\times \boldsymbol{\Omega_{n,\mathbf{k}}}\right], \\
    \dot{\mathbf{k}}&=-e\mathbf{E}+W^2 \text{Im}\left[e\mathbf{E}\times \boldsymbol{\Omega_{n,\mathbf{k}}}\right].
\end{aligned} \right. \label{eom}\end{equation}
Here, $W$ denotes the wave-packet width. The distribution is modified as follows:
\begin{equation} \left\{ \begin{aligned}
    f^{(0)}&=f_0, \\
     f^{(1)}&=\tau e\left[\mathbf{E}-W^2 \mathbf{E}\times \text{Im} \boldsymbol{\Omega^{(0)}_{\mathbf{k}}}\right]\cdot \nabla_{\mathbf{k}}f^{\left(0\right)}, \\
   f^{(2)}&=\tau e\left[\mathbf{E}-W^2 \mathbf{E}\times \text{Im} \boldsymbol{\Omega^{(0)}_{\mathbf{k}}}\right]\cdot \nabla_{\mathbf{k}}f^{\left(1\right)}-\tau W_{\mathbf{k}}^{(2)}\cdot\nabla_{\mathbf{k}}f^{\left(0\right)} \\
   &=\tau e\left[\mathbf{E}-W^2 \mathbf{E}\times \text{Im} \boldsymbol{\Omega^{(0)}_{\mathbf{k}}}\right]\cdot \nabla_{\mathbf{k}}f^{\left(1\right)}-\tau W^2 \text{Im}\left[\nabla_{\mathbf{k}}\xi^{(2)}_{\mathbf{k}}+e\mathbf{E}\times \boldsymbol{\Omega}^{(1)}_{\mathbf{k}}\right]\cdot\nabla_{\mathbf{k}}f^{\left(0\right)},  \\
    ...
\end{aligned} \right. \end{equation}
where $\xi^{(2)}_{\mathbf{k}}=e^2\sum_{\mu,\nu} G^{LR}_{\mu\nu}E^{\mu}E^{\nu}$ and $\Omega_{\alpha}^{(1)}=-2e\epsilon_{\alpha\beta\nu}\partial_{\beta}G_{\mu\nu}^{LR}E^{\mu}$. 

\begin{equation} \left\{ \begin{aligned}
    \mathbf{J}^{(1)}&=-e\int_{\mathbf{k}}f^{(1)}\text{Re}\nabla_{\mathbf{k}}\xi^{(0)}_{\mathbf{k}}+f^{(0)}\text{Re}\left[\nabla_{\mathbf{k}}\xi^{(1)}_{\mathbf{k}}+e\mathbf{E}\times\boldsymbol{\Omega}^{(0)}_\mathbf{k}\right], \\
     &=-e\int_{\mathbf{k}}\tau e\mathbf{E}\cdot \nabla_{\mathbf{k}}f^{\left(0\right)}\text{Re}\nabla_{\mathbf{k}}\xi^{(0)}_{\mathbf{k}}+f^{(0)}e\mathbf{E}\times\text{Re}\boldsymbol{\Omega}^{(0)}_\mathbf{k}-\tau e \left[W^2\left(\mathbf{E}\times\text{Im}\boldsymbol{\Omega}^{(0)}_\mathbf{k}\right)\cdot \nabla_{\mathbf{k}}f^{\left(0\right)}\right] \text{Re}\nabla_{\mathbf{k}}\xi^{(0)}_{\mathbf{k}}, \\
     &=\mathbf{J}^{(1)}\left(W=0\right)+\tau e^2W^2\int_{\mathbf{k}}\left[\left(\mathbf{E}\times\text{Im}\boldsymbol{\Omega}^{(0)}_\mathbf{k}\right)\cdot \nabla_{\mathbf{k}}f^{\left(0\right)}\right] \text{Re}\nabla_{\mathbf{k}}\xi^{(0)}_{\mathbf{k}}, 
\end{aligned} \right. \end{equation}
the linear DC conductivity is now expressed as:
\begin{equation}
    \sigma_{\mu\nu}=\sigma_{\mu\nu}\left(W=0\right)+\tau e^2W^2\int_{\mathbf{k}}\epsilon_{i\nu j}\text{Im}\Omega_{j}^{(0)}\partial_i f_0 \partial_\mu \text{Re}\xi^{(0)}_{\mathbf{k}}
\end{equation}
and
where $\xi^{(2)}_{\mathbf{k}}=e^2\sum_{\mu,\nu} G^{LR}_{\mu\nu}E^{\mu}E^{\nu}$ and $\Omega_{\alpha}^{(1)}=-2e\epsilon_{\alpha\beta\nu}\partial_{\beta}G_{\mu\nu}^{LR}E^{\mu}$. 
\begin{equation} \left\{ \begin{aligned}
    \mathbf{J}^{(2)}&=\mathbf{J}^{(2)}\left(W=0\right)+\tau W^2e^3\int_{\mathbf{k}}\text{Im}\left[\mathbf{E}\times\Omega_{\mathbf{k}}^{(0)}\right]\cdot\nabla_{\mathbf{k}}f_0 \text{Re}\left[\mathbf{E}\times\Omega_{\mathbf{k}}^{(0)}\right]+\tau e \int_{\mathbf{k}}\left\{\text{Im}\left[\nabla_{\mathbf{k}}\xi^{(2)}_{\mathbf{k}}+e\mathbf{E}\times \boldsymbol{\Omega}^{(1)}_{\mathbf{k}}\right]\cdot\nabla_{\mathbf{k}}f_0 \right\}\text{Re}\nabla_{\mathbf{k}}\xi_{\mathbf{k}}^{(0)}\\
    &+\tau^2 e^3\int_{\mathbf{k}}\left\{W^2\mathbf{E}\cdot\nabla_{\mathbf{k}}\left[\text{Im}\left(\mathbf{E}\times\Omega_{\mathbf{k}}^{(0)}\right)\cdot\nabla_{\mathbf{k}}f_0\right]+W^2\text{Im}\left(\mathbf{E}\times\Omega_{\mathbf{k}}^{(0)}\right)\cdot\nabla_{\mathbf{k}}\left[\mathbf{E}\cdot\nabla_{\mathbf{k}}f_0\right]\right\}\text{Re}\nabla_{\mathbf{k}}\xi_{\mathbf{k}}^{(0)} \\
    &-\tau^2 e^3\int_{\mathbf{k}} W^4\text{Im}\left(\mathbf{E}\times\Omega_{\mathbf{k}}^{(0)}\right)\cdot\nabla_{\mathbf{k}}\left[\text{Im}\left(\mathbf{E}\times\Omega_{\mathbf{k}}^{(0)}\right)\cdot\nabla_{\mathbf{k}}f_0\right]\text{Re}\nabla_{\mathbf{k}}\xi_{\mathbf{k}}^{(0)}.
\end{aligned} \right. \end{equation}
The expression for the second-order nonlinear DC conductivity is given by:  

\begin{equation} \begin{aligned}
    \sigma_{\mu\nu\theta}&=\sigma_{\mu\nu\theta}\left(W=0\right)+\tau e^3 W^2\int_{\mathbf{k}}\frac{\epsilon_{i\nu l} \epsilon_{\mu\theta\gamma}\partial_i f_0\text{Im}\Omega_{l}^{(0)}\text{Re}\Omega_{\gamma}^{(0)}+\theta\leftrightarrow \nu}{2}+\frac{\left(2\partial_{\theta}\text{Im}G_{\nu i}^{LR}-\partial_i \text{Im}G_{\nu\theta}^{LR}\right)\partial_{i}f_{0}+\theta\leftrightarrow \nu}{2} \partial_{\mu}\text{Re}\xi_{\mathbf{k}}^{(0)}\\
    &+\tau^2 e^3 W^2\int_{\mathbf{k}}\frac{\epsilon_{j\theta i}\left(\partial_{\nu}\text{Im}\Omega_{i}^{(0)}\partial_j f_0+2\text{Im}\Omega_{i}^{(0)}\partial_j\partial_{\nu}f_0\right)+\theta\leftrightarrow \nu}{2}\partial_{\mu}\text{Re}\xi_{\mathbf{k}}^{(0)} \\
    &+ \tau^2 e^3 W^4 \int_{\mathbf{k}}\frac{\epsilon_{i\nu j}\epsilon_{n\theta\gamma}\text{Im}\Omega^{(0)}_j \partial_i \left( \text{Im}\Omega^{(0)}_{\gamma}\partial_n f_0\right)+\theta\leftrightarrow \nu}{2}\partial_{\mu}\text{Re}\xi_{\mathbf{k}}^{(0)} 
\end{aligned}\end{equation}

In a two-dimensional nonHermitian system, the nonlinear DC conductivity becomes
\begin{equation} \begin{aligned}
    \sigma_{\mu\nu\theta}&=\sigma_{\mu\nu\theta}\left(W=0\right)+\tau e^3 W^2\int_{\mathbf{k}}\frac{\epsilon_{i\nu z} \epsilon_{\mu\theta z}\partial_i f_0\text{Im}\Omega_{z}^{(0)}\text{Re}\Omega_{z}^{(0)}+\theta\leftrightarrow \nu}{2}+\frac{\left(2\partial_{\theta}\text{Im}G_{\nu i}^{LR}-\partial_i \text{Im}G_{\nu\theta}^{LR}\right)\partial_{i}f_{0}+\theta\leftrightarrow \nu}{2} \partial_{\mu}\text{Re}\xi_{\mathbf{k}}^{(0)}\\
    &+\tau^2 e^3 W^2\int_{\mathbf{k}}\frac{\epsilon_{j\theta z}\left(\partial_{\nu}\text{Im}\Omega_{z}^{(0)}\partial_j f_0+2\text{Im}\Omega_{z}^{(0)}\partial_j\partial_{\nu}f_0\right)+\theta\leftrightarrow \nu}{2}\partial_{\mu}\text{Re}\xi_{\mathbf{k}}^{(0)} \\
    &+ \tau^2 e^3 W^4 \int_{\mathbf{k}}\frac{\epsilon_{i\nu z}\epsilon_{n\theta z}\text{Im}\Omega^{(0)}_z \partial_i \left( \text{Im}\Omega^{(0)}_{z}\partial_n f_0\right)+\theta\leftrightarrow \nu}{2}\partial_{\mu}\text{Re}\xi_{\mathbf{k}}^{(0)} 
\end{aligned}\end{equation}
Specially,
\begin{equation} \begin{aligned}
    \sigma_{\mu\mu\mu}&=\sigma_{\mu\mu\mu}\left(W=0\right)+\tau e^3 W^2\int_{\mathbf{k}}\left(2\partial_{\mu}\text{Im}G_{\mu i}^{LR}-\partial_i \text{Im}G_{\mu\mu}^{LR}\right)\partial_{i}f_{0}\partial_{\mu}\text{Re}\xi_{\mathbf{k}}^{(0)}\\
    &+\tau^2 e^3 W^2\int_{\mathbf{k}}\epsilon_{j\mu z}\left(\partial_{\mu}\text{Im}\Omega_{z}^{(0)}\partial_j f_0+2\text{Im}\Omega_{z}^{(0)}\partial_j\partial_{\mu}f_0\right)\partial_{\mu}\text{Re}\xi_{\mathbf{k}}^{(0)} \\
    &+ \tau^2 e^3 W^4 \int_{\mathbf{k}}\epsilon_{i\mu z}\epsilon_{n\mu z}\text{Im}\Omega^{(0)}_z \partial_i \left( \text{Im}\Omega^{(0)}_{z}\partial_n f_0\right)\partial_{\mu}\text{Re}\xi_{\mathbf{k}}^{(0)} 
\end{aligned}\end{equation}

\begin{equation} \begin{aligned}
    \sigma_{xxx}&=\sigma_{xxx}\left(W=0\right)+\tau e^3 W^2\sum_{i\in{x,y}}\int_{\mathbf{k}}\left(2\partial_{x}\text{Im}G_{x i}^{LR}-\partial_i \text{Im}G_{xx}^{LR}\right)\partial_{i}f_{0}\partial_{x}\text{Re}\xi_{\mathbf{k}}^{(0)}\\
    &-\tau^2 e^3 W^2\int_{\mathbf{k}}\left(\partial_{x}\text{Im}\Omega_{z}^{(0)}\partial_y f_0+2\text{Im}\Omega_{z}^{(0)}\partial_y\partial_{x}f_0\right)\partial_{x}\text{Re}\xi_{\mathbf{k}}^{(0)} \\
    &+ \tau^2 e^3 W^4 \int_{\mathbf{k}}\text{Im}\Omega^{(0)}_z \partial_y \left( \text{Im}\Omega^{(0)}_{z}\partial_y f_0\right)\partial_{x}\text{Re}\xi_{\mathbf{k}}^{(0)} 
\end{aligned}\end{equation}
and 

\begin{equation} \begin{aligned}
    \sigma_{yyy}&=\sigma_{yyy}\left(W=0\right)+\tau e^3 W^2\sum_{i\in{x,y}}\int_{\mathbf{k}}\left(2\partial_{y}\text{Im}G_{y i}^{LR}-\partial_i \text{Im}G_{yy}^{LR}\right)\partial_{i}f_{0}\partial_{y}\text{Re}\xi_{\mathbf{k}}^{(0)}\\
    &+\tau^2 e^3 W^2\int_{\mathbf{k}}\left(\partial_{y}\text{Im}\Omega_{z}^{(0)}\partial_x f_0+2\text{Im}\Omega_{z}^{(0)}\partial_x\partial_{y}f_0\right)\partial_{y}\text{Re}\xi_{\mathbf{k}}^{(0)} \\
    &+ \tau^2 e^3 W^4 \int_{\mathbf{k}}\text{Im}\Omega^{(0)}_z \partial_x \left( \text{Im}\Omega^{(0)}_{z}\partial_x f_0\right)\partial_{y}\text{Re}\xi_{\mathbf{k}}^{(0)} 
\end{aligned}\end{equation}

\begin{equation} \begin{aligned}
    \sigma_{xyy}&=\sigma_{xyy}\left(W=0\right)+\tau e^3 W^2\int_{\mathbf{k}}\partial_x f_0\text{Im}\Omega_{z}^{(0)}\text{Re}\Omega_{z}^{(0)}+\sum_{i\in{x,y}}\left(2\partial_{y}\text{Im}G_{y i}^{LR}-\partial_i \text{Im}G_{yy}^{LR}\right)\partial_{i}f_{0} \partial_{x}\text{Re}\xi_{\mathbf{k}}^{(0)}\\
    &+\tau^2 e^3 W^2\int_{\mathbf{k}}\left(\partial_{y}\text{Im}\Omega_{z}^{(0)}\partial_x f_0+2\text{Im}\Omega_{z}^{(0)}\partial_j\partial_{y}f_0\right)\partial_{x}\text{Re}\xi_{\mathbf{k}}^{(0)} \\
    &+ \tau^2 e^3 W^4 \int_{\mathbf{k}}\text{Im}\Omega_{z}^{(0)} \partial_x \left( \text{Im}\Omega_{z}^{(0)}\partial_x f_0\right)\partial_{x}\text{Re}\xi_{\mathbf{k}}^{(0)} 
\end{aligned}\end{equation}

\end{document}